\documentclass[final,5p,times,twocolumn]{elsarticle}

\usepackage{pifont}
\usepackage{color}
\usepackage{amsmath}
\usepackage{amssymb}
\usepackage{graphicx}
\usepackage{lineno}

\usepackage{type1cm}
\usepackage{eso-pic}

\newcommand*\patchAmsMathEnvironmentForLineno[1]{%
  \expandafter\let\csname old#1\expandafter\endcsname\csname #1\endcsname
  \expandafter\let\csname oldend#1\expandafter\endcsname\csname end#1\endcsname
  \renewenvironment{#1}%
     {\linenomath\csname old#1\endcsname}%
     {\csname oldend#1\endcsname\endlinenomath}}%
\newcommand*\patchBothAmsMathEnvironmentsForLineno[1]{%
  \patchAmsMathEnvironmentForLineno{#1}%
  \patchAmsMathEnvironmentForLineno{#1*}}%
\AtBeginDocument{%
\patchBothAmsMathEnvironmentsForLineno{equation}%
\patchBothAmsMathEnvironmentsForLineno{align}%
\patchBothAmsMathEnvironmentsForLineno{flalign}%
\patchBothAmsMathEnvironmentsForLineno{alignat}%
\patchBothAmsMathEnvironmentsForLineno{gather}%
\patchBothAmsMathEnvironmentsForLineno{multline}%
}

\makeatletter
\def\ps@pprintTitle{%
 \let\@oddhead\@empty
 \let\@evenhead\@empty
 \def\@oddfoot{}%
 \let\@evenfoot\@oddfoot}
\makeatother

\usepackage{hyperref}
\hypersetup{
 colorlinks=true,     
 urlcolor=blue,       
 linkcolor=blue,     
 citecolor=blue,     
 plainpages=false,
 breaklinks=true,
 bookmarksnumbered=true,
 bookmarksopen=true,
 pdfkeywords={},
 pdfpagemode=UseOutlines  
 }

\journal{}

\begin{document}

\title{Precision Measurements of $A_1^n$ in the Deep Inelastic Regime}

\author[cmu,cenpa]{D.~S.~Parno\corref{cor1}}
\author[temple,umass]{D.~Flay}
\author[temple]{M.~Posik}
\author[uky]{K.~Allada}
\author[temple]{W.~Armstrong}
\author[wm]{T.~Averett}
\author[cmu]{F.~Benmokhtar}
\author[mit]{W.~Bertozzi}
\author[jlab]{A.~Camsonne}
\author[odu]{M.~Canan}
\author[uva]{G.~D.~Cates}
\author[hampton]{C.~Chen}
\author[jlab]{J.-P.~Chen}
\author[seoul]{S.~Choi}
\author[jlab]{E.~Chudakov}
\author[infn-rome,sanita]{F.~Cusanno\corref{deceased}}
\author[uva]{M.~M.~Dalton}
\author[mit]{W.~Deconinck}
\author[jlab,uva]{C.~W.~de~Jager}
\author[uva]{X.~Deng}
\author[jlab]{A.~Deur}
\author[uky]{C.~Dutta}
\author[odu,rutgers]{L.~El~Fassi}
\author[cmu]{G.~B.~Franklin}
\author[cmu]{M.~Friend}
\author[duke]{H.~Gao}
\author[infn-rome]{F.~Garibaldi}
\author[mit]{S.~Gilad}
\author[jlab,rutgers]{R.~Gilman}
\author[kharkov]{O.~Glamazdin}
\author[odu]{S.~Golge}
\author[jlab]{J.~Gomez}
\author[lanl]{L.~Guo}
\author[jlab]{O.~Hansen}
\author[jlab]{D.~W.~Higinbotham}
\author[longwood]{T.~Holmstrom}
\author[mit]{J.~Huang}
\author[odu,pascal]{C.~Hyde}
\author[cairo]{H.~F.~Ibrahim}
\author[rutgers,lanl]{X.~Jiang}
\author[uva]{G.~Jin}
\author[wm]{J.~Katich}
\author[wm]{A.~Kelleher}
\author[uky]{A.~Kolarkar}
\author[uky]{W.~Korsch}
\author[rutgers]{G.~Kumbartzki}
\author[jlab]{J.J.~LeRose}
\author[uva]{R.~Lindgren}
\author[uva]{N.~Liyanage}
\author[kent]{E.~Long}
\author[temple]{A.~Lukhanin}
\author[cmu]{V.~Mamyan}
\author[umass]{D.~McNulty}
\author[temple]{Z.-E.~Meziani}
\author[jlab]{R.~Michaels}
\author[stefan]{M.~Mihovilovi\v{c}}
\author[mit,jlab]{B.~Moffit}
\author[mit]{N.~Muangma}
\author[jlab]{S.~Nanda}
\author[missu]{A.~Narayan}
\author[uva]{V.~Nelyubin}
\author[uva]{B.~Norum}
\author[missu]{Nuruzzaman}
\author[seoul]{Y.~Oh}
\author[uiuc]{J.~C.~Peng}
\author[duke,caltech]{X.~Qian} 
\author[duke,jlab]{Y.~Qiang}
\author[syracuse]{A.~Rakhman}
\author[uva,umass]{S.~Riordan}
\author[jlab]{A.~Saha\corref{deceased}} 
\author[temple,jlab]{B.~Sawatzky}
\author[uva]{M.~H.~Shabestari}
\author[yerevan]{A.~Shahinyan}
\author[ljubljana,stefan]{S.~\v{S}irca}
\author[anl,jlab]{P.~Solvignon}
\author[uva]{R.~Subedi}
\author[mit,jlab]{V.~Sulkosky}
\author[uva]{W.~A.~Tobias}
\author[longwood]{W.~Troth}
\author[uva]{D.~Wang}  
\author[uiuc]{Y.~Wang}
\author[jlab]{B.~Wojtsekhowski}
\author[ustc]{X.~Yan}
\author[temple,wm]{H.~Yao}
\author[ustc]{Y.~Ye}
\author[hampton]{Z.~Ye}
\author[hampton]{L.~Yuan}
\author[mit]{X.~Zhan}
\author[lanzhou]{Y.~Zhang} 
\author[lanzhou,rutgers]{Y.-W.~Zhang} 
\author[wm]{B.~Zhao}
\author[uva]{X.~Zheng}
\author{(The Jefferson Lab Hall A Collaboration)}

\address[cmu]{Carnegie Mellon University, Pittsburgh, PA 15213}
\address[cenpa]{Center for Experimental Nuclear Physics and Astrophysics and Department of Physics, University of Washington, Seattle, WA 98195}
\address[temple]{Temple University, Philadelphia, PA 19122}
\address[umass]{University of Massachusetts, Amherst, MA 01003}
\address[uky]{University of Kentucky, Lexington, KY 40506}
\address[wm]{College of William and Mary, Williamsburg, VA 23187}
\address[mit]{Massachusetts Institute of Technology, Cambridge, MA 02139}
\address[jlab]{Thomas Jefferson National Accelerator Facility, Newport News, VA 23606}
\address[odu]{Old Dominion University, Norfolk, VA 23529}
\address[uva]{University of Virginia, Charlottesville, VA 22904}
\address[hampton]{Hampton University, Hampton, VA 23187}
\address[seoul]{Seoul National University, Seoul 151-742, South Korea}
\address[infn-rome]{INFN, Sezione di Roma, I-00161 Rome, Italy}
\address[sanita]{Istituto Superiore di Sanit\`a, I-00161 Rome, Italy}
\address[rutgers]{Rutgers, The State University of New Jersey, Piscataway, NJ 08855}
\address[duke]{Duke University, Durham, NC 27708}
\address[kharkov]{Kharkov Institute of Physics and Technology, Kharkov 61108, Ukraine}
\address[lanl]{Los Alamos National Laboratory, Los Alamos, NM 87545}
\address[longwood]{Longwood University, Farmville, VA 23909}
\address[pascal]{Universit\'e Blaise Pascal/IN2P3, F-63177 Aubi\`ere, France}
\address[cairo]{Cairo University, Giza 12613, Egypt}
\address[kent]{Kent State University, Kent, OH 44242}
\address[stefan]{Jo\v{z}ef Stefan Institute, Ljubljana, Slovenia}
\address[missu]{Mississippi State University, MS 39762}
\address[uiuc]{University of Illinois at Urbana-Champaign, Urbana, IL 61801}
\address[caltech]{Kellogg Radiation Laboratory, California Institute of Technology, Pasadena, CA 91125}
\address[syracuse]{Syracuse University, Syracuse, NY 13244}
\address[yerevan]{Yerevan Physics Institute, Yerevan 375036, Armenia}
\address[ljubljana]{University of Ljubljana, SI-1000 Ljubljana, Slovenia}
\address[anl]{Argonne National Lab, Argonne, IL 60439}
\address[ustc]{University of Science and Technology of China, Hefei 230026, People's Republic of China}
\address[lanzhou]{Lanzhou University, Lanzhou 730000, Gansu, People's Republic of China}

\cortext[cor_dsp]{
E-mail: dparno@uw.edu}
\cortext[deceased]{Deceased}

\date{\today}

\begin{abstract}
We have performed precision measurements of the double-spin virtual-photon asymmetry $A_1$ on the neutron in the deep inelastic scattering regime, using an open-geometry, large-acceptance spectrometer. Our data cover a wide kinematic range $0.277 \leq x \leq 0.548$ at an average $Q^2$ value of 3.078~(GeV/c)$^2$, doubling the available high-precision neutron data in this $x$ range. We have combined our results with world data on proton targets to extract the ratio of polarized-to-unpolarized parton distribution functions for up quarks and for down quarks in the same kinematic range. Our data are consistent with a previous observation of an $A_1^n$ zero crossing near $x=0.5$. We find no evidence of a transition to a positive slope in $(\Delta d + \Delta \bar{d})/(d + \bar{d})$ up to $x=0.548$. 
\end{abstract}


\maketitle


Ever since the European Muon Collaboration determined that the quark-spin contribution was insufficient to account for the spin of the proton~\cite{Ashman:EMC89}, the origin of the nucleon spin has been an open puzzle; see Ref.~\cite{aidala:2013} for a recent review. Recently, studies of polarized proton-proton collisions have found evidence for a non-zero contribution from the gluon spin~\cite{deFlorian:2014} and for a significantly positive polarization of $\bar{u}$ quarks~\cite{adamczyk:2014}. The possible contribution of parton orbital angular momentum (OAM) is also under investigation. In the valence quark region, combining spin-structure data obtained in polarized-lepton scattering on protons and neutrons allows the separation of contributions from up and down quarks and permits a sensitive test of several theoretical models.

In deep inelastic scattering (DIS), nucleon structure is conventionally parameterized by the unpolarized structure functions $F_1(x, Q^2)$ and $F_2(x, Q^2)$, and by the polarized structure functions $g_1(x, Q^2)$ and $g_2(x, Q^2)$, where $Q^2$ is the negative square of the four-momentum transferred in the scattering interaction and $x$ is the Bjorken scaling variable, which at leading order in the infinite-momentum frame equals the fraction of the nucleon momentum carried by the struck quark. One useful probe of the nucleon spin structure is the asymmetry $A_1 = (\sigma_{1/2} - \sigma_{3/2})/(\sigma_{1/2} + \sigma_{3/2})$, where $\sigma_{1/2 (3/2)}$ is the cross section of virtual photoabsorption on the nucleon for a total spin projection of $1/2$ ($3/2$) along the virtual-photon momentum direction. At finite $Q^2$, this asymmetry may be expressed in terms of the nucleon structure functions as~\cite{Melnitchouk:BackgroundQHD04}
\begin{equation}
\label{eq:def_A1}
A_1 (x, Q^2) = \left[ g_1(x, Q^2) - \gamma^2 g_2(x, Q^2) \right] / F_1(x, Q^2),
\end{equation}

\noindent where $\gamma^2 = 4 M^2 x^2 c^2 / Q^2$ and $M$ is the nucleon mass. For large $Q^2$, $\gamma^2 \ll 1$ and $A_1 (x) \approx g_1(x)/F_1(x)$; since $g_1$ and $F_1$ have the same $Q^2$ evolution to leading order~\cite{dokshitser:1977, gribov:1972, altarelli:1977}, $A_1$ may be approximated as a function of $x$ alone. Through Eq.~\ref{eq:def_A1}, measurements of $A_1$ on proton and neutron targets also allow extraction of the flavor-separated ratios of polarized to unpolarized parton distribution functions (PDFs), $(\Delta q(x) + \Delta \bar{q}(x))/ (q(x)+ \bar{q}(x))$. Here, $q(x) = q^{\uparrow}(x) + q^{\downarrow}(x)$ and $\Delta q(x) = q^{\uparrow}(x) - q^{\downarrow}(x)$, where $q^{\uparrow(\downarrow)}(x)$ is the probability of finding the quark $q$ with a given value of $x$ and with spin (anti)parallel to that of the nucleon. This Letter reports a high-precision measurement of the neutron $A_1$, $A_1^n$, in a kinematic range where theoretical predictions begin to diverge.

A variety of theoretical approaches predict that $A_1^{n} \rightarrow 1$ as $x \rightarrow 1$. Calculations in the relativistic constituent quark model (RCQM), for example, generally assume that SU(6) symmetry is broken via a color hyperfine interaction between quarks, lowering the energy of spectator-quark pairs in a spin singlet state relative to those in a spin triplet state and increasing the probability that, at high $x$, the struck quark carries the nucleon spin~\cite{Isgur:SU6Hyperfine99}. 

In perturbative quantum chromodynamics (pQCD), valid at large $x$ and large $Q^2$ where the coupling of gluons to the struck quark is small, the leading-order assumption that the valence quarks have no OAM leads to the same conclusion about the spin of the struck quark~\cite{Farrar:HHC75, Farrar:HHC77}. Parameterizations of the world data, in the context of pQCD models, have been made at next to leading order (NLO) both with and without this assumption of hadron helicity conservation. The LSS(BBS) parameterization~\cite{Leader:LSSBBS} is a classic example of the former; Avakian~\textit{et al.}~\cite{Avakian:OAM07} later extended that parameterization to explicitly include Fock states with nonzero quark OAM. Both parameterizations enforce $A_1^n(x\rightarrow0) < 0$ and $A_1^n(x\rightarrow1) \rightarrow 1$ and predict $\lim_{x \to 1}(\Delta d + \Delta \bar{d})/(d + \bar{d}) = 1$, but the OAM-inclusive parameterization predicts a zero crossing at significantly higher $x$. Recently, the Jefferson Lab Angular Momentum (JAM) collaboration performed a global NLO analysis at $Q^2 = 1$~(GeV/c)$^2$ to produce a new parameterization~\cite{jimenez-delgado:2014a}, and then systematically studied the effects of various input assumptions~\cite{jimenez-delgado:2014b}. Without enforcing hadron helicity conservation, JAM found that the ratio $(\Delta d + \Delta \bar{d})/(d + \bar{d})$ remains negative across all $x$; regardless of this initial assumption, the existing world data can be fit approximately equally well with or without explicit OAM terms in the form given by Ref.~\cite{Avakian:OAM07}. The scarcity of precise DIS neutron data above $x \approx 0.4$, combined with the absence of such data points for $x \gtrsim 0.6$, leaves the pQCD parameterizations remarkably unconstrained.

The statistical model treats the nucleon as a gas of massless partons at thermal equilibrium, using both chirality and DIS data to constrain the thermodynamical potential of each parton species. At a moderate $Q^2$ value of 4~(GeV/c)$^2$, $A_1^{n}(x \rightarrow 1) \rightarrow 0.6 \cdot \Delta u(x)/u(x) \sim 0.46$~\cite{Bourrely:Stats02}. Statistical-model predictions are thus in conflict with hadron helicity conservation. A modified Nambu-Jona-Lasinio (NJL) model, including both scalar and axial-vector diquark channels, yields a similar prediction for $A_1^n$ as $x \rightarrow 1$~\cite{cloet:2005}. A recent approach based on Dyson-Schwinger equations (DSE) predicts $A_1^n(x=1) = 0.34$ in a contact-interaction framework, and 0.17 in a more realistic framework in which the dressed-quark mass is permitted to depend on momentum~\cite{roberts:2013}; the latter prediction is significantly smaller than either the statistical or NJL prediction at $x=1$. However, existing DIS data do not extend to high enough $x$ to definitively favor one model over another.

\begin{table*}[hbt]
\begin{center}
\caption{DIS asymmetries $A_\parallel$ and $A_\perp$ measured on $^3$He at two beam energies.}
\label{tab:results_3He_asy}
\begin{tabular}{cccrr}
\multicolumn{1}{c}{$E$ (GeV)} & \multicolumn{1}{c}{$\langle x \rangle$} & \multicolumn{1}{c}{$\langle Q^2 \rangle$ (GeV/c)$^2$} & \multicolumn{1}{c}{$A_\parallel \pm \textrm{stat} \pm \textrm{syst}$} & \multicolumn{1}{c}{$A_\perp \pm \textrm{stat} \pm \textrm{syst}$}  \\
\hline
4.74 	& 0.277 & 2.038 & $-0.008 \pm 0.015 \pm 0.007$ & $-0.002 \pm 0.008 \pm 0.003$ \\
		& 0.325 & 2.347 & $-0.009 \pm 0.009 \pm 0.003$ & $-0.001 \pm 0.005 \pm 0.002$ \\
		& 0.374 & 2.639 & $0.005 \pm 0.007 \pm 0.002$ & $-0.011 \pm 0.004 \pm 0.002$ \\
		& 0.424 & 2.915 & $-0.025 \pm 0.007 \pm 0.005$ & $-0.003 \pm 0.004 \pm 0.002$ \\
		& 0.473 & 3.176 & $-0.021 \pm 0.008 \pm 0.003$ & $-0.005 \pm 0.004 \pm 0.001$ \\
\hline
5.89 & 0.277 & 2.626 & $0.019 \pm 0.027 \pm 0.010$ & $0.010 \pm 0.008 \pm 0.003$ \\
		& 0.325 & 3.032 & $-0.017 \pm 0.012 \pm 0.003$ & $0.004 \pm 0.004 \pm 0.001$ \\
		& 0.374 & 3.421 & $-0.006 \pm 0.009 \pm 0.002$ & $-0.001 \pm 0.003 \pm 0.001$ \\
		& 0.424 & 3.802 & $-0.020 \pm 0.009 \pm 0.003$ & $-0.004 \pm 0.003 \pm 0.001$ \\
		& 0.474 & 4.169 & $-0.021 \pm 0.010 \pm 0.006$ & $0.000 \pm 0.003 \pm 0.001$ \\
		& 0.524 & 4.514 & $0.002 \pm 0.012 \pm 0.002$ & $0.000 \pm 0.004 \pm 0.001$ \\
		& 0.573 & 4.848 & $0.003 \pm 0.015 \pm 0.003$ & $0.003 \pm 0.004 \pm 0.001$ \\
\end{tabular}
\end{center}
\label{default}
\end{table*}%

Measurements of the virtual photon asymmetry $A_1$ can be made via doubly polarized electron-nucleon scattering. With both beam and target polarized longitudinally with respect to the beamline, $A_{\parallel} = (\sigma^{\downarrow \Uparrow} - \sigma^{\uparrow \Uparrow})/(\sigma^{\downarrow \Uparrow} + \sigma^{\uparrow \Uparrow})$ is the scattering asymmetry between configurations with the electron spin anti-aligned ($\downarrow$) and aligned ($\uparrow$) with the beam direction. Meanwhile, $A_{\perp} = (\sigma^{\downarrow \Rightarrow} - \sigma^{\uparrow \Rightarrow})/(\sigma^{\downarrow \Rightarrow} + \sigma^{\uparrow \Rightarrow})$ is measured with the target spin lying in
the nominal scattering plane, perpendicular to the incident beam direction and on the side of the scattered electron. $A_1$ may be related to these asymmetries through~\cite{Melnitchouk:BackgroundQHD04}:
\begin{equation}
A_1 = \frac{1}{D \left( 1 + \eta \xi \right)} A_{\parallel} - \frac{\eta}{d \left( 1 + \eta \xi \right)} A_{\perp},
\end{equation}

\noindent where the kinematic variables are given in the laboratory frame by $D = (E - \epsilon E')/(E (1 + \epsilon R))$, $\eta = \epsilon \sqrt{Q^2}/(E - \epsilon E')$, $d = D \sqrt{2\epsilon/(1 + \epsilon)}$, and $\xi = \eta (1+ \epsilon)/2\epsilon$. Here, $E$ is the initial electron energy; $E'$ is the scattered electron energy; $\epsilon = 1/[1 + 2 (1 + 1/\gamma^2)\tan^2(\theta/2)]$; $\theta$ is the electron scattering angle; and $R = \sigma_L/\sigma_T$, parameterized via R1998~\cite{Abe:e143_99}, is the ratio of the longitudinal to the transverse virtual photoabsorption cross sections. 

Experiment E06-014 ran in Hall A of Jefferson Lab in February and March 2009 with the primary purpose of measuring a twist-3 matrix element of the neutron~\cite{posik:2014}. Longitudinally polarized electrons were generated via illumination of a strained superlattice GaAs photocathode by circularly polarized laser light~\cite{Sinclair:PolarizedSource07} and delivered to the experimental hall with energies of 4.7~and 5.9~GeV. The rastered 12-15 $\mu$A beam was incident on a target of $^3$He gas, polarized in the longitudinal and transverse directions via spin-exchange optical pumping of a Rb-K mixture~\cite{Babcock:HSEOP03} and contained in a 40-cm-long glass cell. The left high-resolution spectrometer~\cite{Alcorn:HallA04} and BigBite spectrometer~\cite{deLange:BBgeneralNIKHEF} independently detected scattered electrons at angles of 45$^{\circ}$ on beam left and right, respectively. 

The longitudinal beam polarization was monitored continuously by Compton polarimetry~\cite{Escoffier:ComptonEBP05, friend:2012} and intermittently by M{\o}ller polarimetry~\cite{Glamazdin:Moller1999}. In three run periods with polarized beam, the longitudinal beam polarization $P_b$ averaged $0.74 \pm 0.01$ ($E=5.9$~GeV), $0.79 \pm 0.01$ ($E=5.9$~GeV), and $0.63 \pm 0.01$ ($E=4.7$~GeV). A feedback loop limited the charge asymmetry to within 100~ppm. The target polarization $P_t$, averaging about 50\%, was measured periodically using nuclear magnetic resonance~\cite{romalis:1998} and calibrated with electron paramagnetic resonance; in the longitudinal orientation, the calibration was cross-checked with nuclear magnetic resonance data from a well-understood water target. 

The raw asymmetry $A^{\mathrm{raw}}_{\parallel(\perp)}$ was corrected for beam and target effects according to $A^{\mathrm{cor}}_{\parallel(\perp)} = A^{\mathrm{raw}}_{\parallel(\perp)}/[P_b P_t f_{\mathrm{N}_2}(\cos\phi)]$, where the dilution factor $f_{\mathrm{N}_2}$, determined from dedicated measurements with a nitrogen target, corrects for scattering from the small amount of N$_2$ gas added to the $^3$He target to reduce depolarization effects~\cite{Kominis:th01}. The angle $\phi$, which appears in $A^{\mathrm{cor}}_\perp$, lies between the scattering plane, defined by the initial and final electron momenta, and the polarization plane, defined by the electron and target spins.

Data for the asymmetry measurements were taken with the BigBite detector stack, which in this configuration included eighteen wire planes in three orientations, a gas \v{C}erenkov detector~\cite{posik:phd2014}, a pre-shower $+$ shower calorimeter, and a scintillator plane between the calorimeter layers. The primary trigger was formed when signals above threshold were registered in geometrically overlapping regions of the gas \v{C}erenkov and calorimeter. Wire-plane data allowed momentum reconstruction with a resolution of 1\%~\cite{posik:phd2014}. With an angular acceptance of 65~msr, BigBite continuously measured electrons over the entire kinematic range of the experiment, and the sample was later divided into $x$ bins of equal size. 

Pair-produced electrons, originating from $\pi^0$ decay, contaminate the sample of DIS electrons, especially in the lowest $x$ bins. We measured the yield of this process by reversing the BigBite polarity to observe $e^+$ with the same acceptance. A fit to these data, combined with data from the left high-resolution spectrometer and with CLAS EG1b~\cite{Dharmawardane:CLASA1meas06} data taken at a similar scattering angle, was used to fill gaps in the kinematic coverage of these special measurements. The resulting ratio $f_{e^+} = N_{e^+}/N_{e^-}$ quantifies the contamination of the electron sample with pair-produced electrons. The underlying double-spin asymmetry $A^{e^+}$ of the $\pi^0$ production process was measured to be $1-2\%$ using the positron sample obtained during normal BigBite running, and cross-checked against the reversed-polarity positron asymmetry for the available kinematics.

The contamination of the scattered-electron sample with $\pi^-$ was below 3\% in all $x$ bins, limited primarily by the efficiency of the gas \v{C}erenkov in eliminating pions from the online trigger. Due to the low contamination level, the asymmetry in pion production had a negligible ($\lesssim 1\%$) effect on $A_\parallel$ and $A_\perp$, and the pion correction to the asymmetry was therefore treated as a pure dilution $f_{\pi^-}$. Contamination of the positron sample with $\pi^+$ resulted in the dilution factor $f_{\pi^+}$. Particle identification was the dominant overall source of systematic error in this measurement.

The final physics asymmetries $A_{\parallel(\perp)}$, which are listed in Table~\ref{tab:results_3He_asy}, include internal and external radiative corrections $\Delta A_{\parallel(\perp)}^{RC}$ as well as background corrections: 
\begin{equation}
A_{\parallel(\perp)} = \frac{A^{\mathrm{cor}}_{\parallel(\perp)} - f_{e^+} A^{e^+}_{\parallel(\perp)}}{1- f_{\pi^-} - f_{e^+} + f_{\pi^+} f_{e^+}} + \Delta A_{\parallel(\perp)}^{RC}.
\end{equation} 

\noindent To compute $\Delta A_{\parallel(\perp)}^{RC}$, the asymmetries were reformulated as polarized cross-section differences using the F1F209~\cite{f1f209} parameterization for the radiated unpolarized cross section. The polarized elastic tail was computed~\cite{amroun:1994} and found to be negligible in both the parallel and perpendicular cases; therefore, this tail was not subtracted.  Radiative corrections were then applied iteratively, according to the formalism first described by Mo and Tsai~\cite{Mo:EnergyLoss69, TsaiSLAC} for the unpolarized case, and checked by the Akushevich et al.~\cite{polrad:1997} formalism for the polarized case. The DSSV model~\cite{deFlorian:DSSV2008} was used as an input for the DIS region; the integration phase space was completed in the resonance region with the MAID model~\cite{maid:2007}, and in the quasi-elastic region with the Bosted nucleon form factors~\cite{bosted:1995} smeared with a scaling function~\cite{amaro:2005}. The final results were then converted back to asymmetries. The contribution of these corrections to the uncertainty on $A_{\parallel(\perp)}$, estimated by varying the input models and radiation thicknesses of materials in the beamline and along the trajectory of the scattered electrons, was $\lesssim 2\%$. Smearing effects across individual $x$ bins, due to the finite detector resolution, contributed a negligible amount to this error. Energy-loss calculations were performed within the radiative-correction framework and not as part of the acceptance calculation.

Polarized $^3$He targets are commonly used as effective polarized neutron targets because, in the dominant $S$ state, the spin of the $^3$He nucleus is carried by the neutron. To extract the neutron asymmetry $A_1^n$ from the measured asymmetry $A_1^{^3\mathrm{He}}$ on the nuclear target, we used a model for the $^3$He wavefunction incorporating $S$, $S'$, and $D$ states as well as a pre-existing $\Delta(1232)$ component~\cite{Bissey:NuclearEffects2002}:
\begin{equation}
A_1^n = \frac{F_2^{^3\mathrm{He}} \left[ A_1^{^3\mathrm{He}} - 2 \frac{F_2^p}{F_2^{^3\mathrm{He}}} P_p A_1^p \left( 1 - \frac{0.014}{2P_p} \right) \right]}{P_n F_2^n \left( 1 + \frac{0.056}{P_n} \right)}.
\end{equation}

\begin{figure}[tbp]
\begin{center}
        \includegraphics[width=\columnwidth]{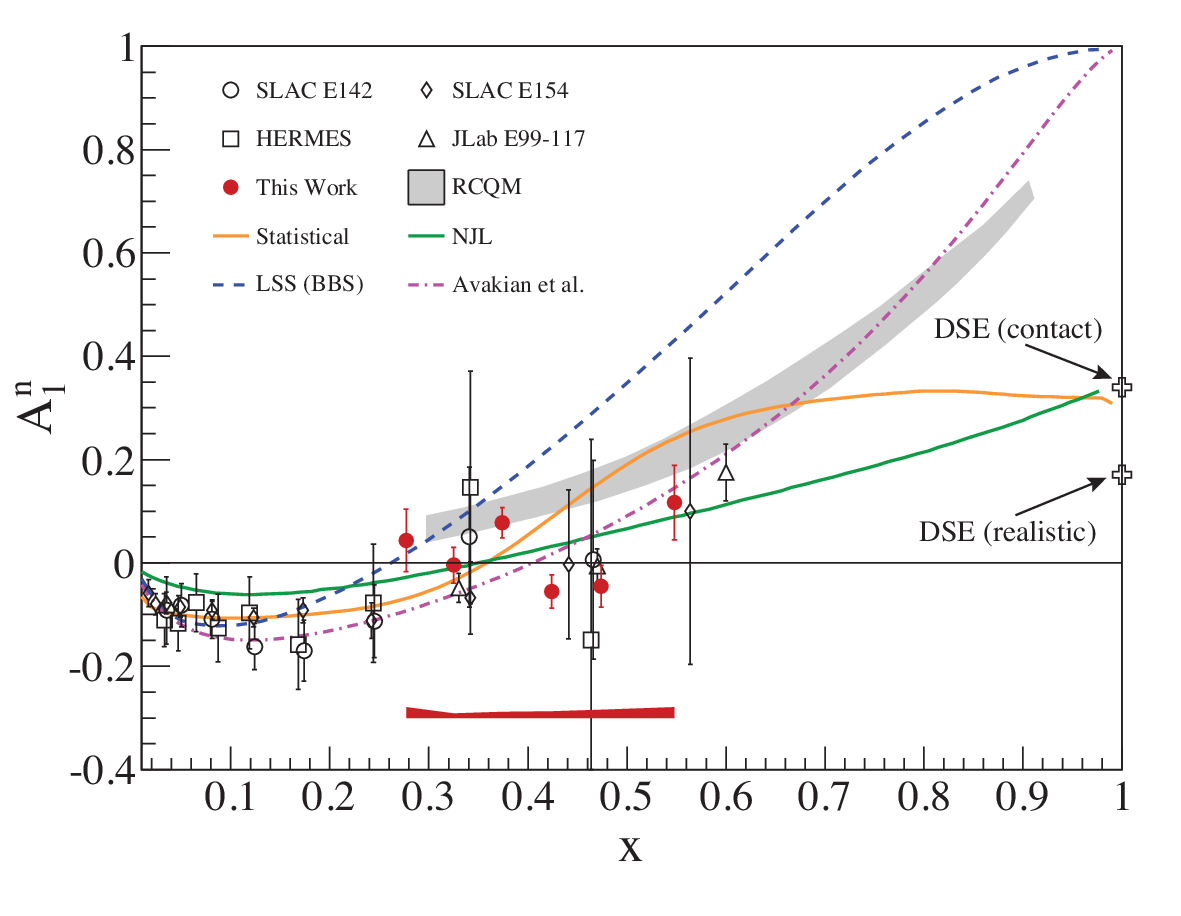}
        \caption{(Color online) Our $A_1^n$ results in the DIS regime (filled circles), compared with world $A_1^n$ data extracted using $^3$He targets (SLAC E142~\cite{Anthony:e142_meas96}, SLAC E154~\cite{Abe:e154_97}, Jefferson Lab E99-117~\cite{zheng:2004a}, and HERMES~\cite{ackerstaff:1997}). Statistical uncertainties are shown as error bars; our systematic uncertainties are given by the band below the data. Selected model predictions are also shown: RCQM~\cite{Isgur:SU6Hyperfine99}, statistical~\cite{Bourrely:Stats02, statmodel_updated_2014}, NJL~\cite{cloet:2005}, and two DSE-based approaches~\cite{roberts:2013} (crosses at $x=1$). Quark OAM is assumed to be absent in the LSS(BBS) parameterization~\cite{Leader:LSSBBS}, but is explicitly allowed in the Avakian~\textit{et al.} parameterization~\cite{Avakian:OAM07}. The recent pQCD parameterizations from the JAM collaboration were performed at $Q^2 \approx 1$~(GeV/c)$^2$ and are not plotted with our higher-$Q^2$ data.}
        \label{fig:a1n}
\end{center}
\end{figure}

\noindent The effective proton and neutron polarizations were taken as $P_p = -0.028^{+0.009}_{-0.004}$ and $P_n = 0.860^{+0.036}_{-0.020}$~\cite{zheng:2004}. $F_2$ was parameterized with F1F209~\cite{f1f209} for $^3$He and with CJ12~\cite{cj12} for the neutron and proton, while $A_1^p$ was modeled with a $Q^2$-independent, three-parameter fit to world data~\cite{adeva:1998, airapetian:2007, ashman:1988, Ashman:EMC89, Abe:e143_98, anthony:1999, Dharmawardane:CLASA1meas06} on proton targets. Corrections were applied separately to the two beam energies, at the average measured $Q^2$ values of 2.59~(GeV/c)$^2$ ($E=4.7$~GeV) and 3.67~(GeV/c)$^2$ ($E=5.9$~GeV). The resulting neutron asymmetry, the statistics-weighted average of the asymmetries measured at the two beam energies, is given as a function of $x$ in Fig.~\ref{fig:a1n} and Table~\ref{tab:results_a1} and corresponds to an average $Q^2$ value of 3.078~(GeV/c)$^2$. Table~\ref{tab:results_a1} also gives our results for the structure-function ratio $g_1^n/F_1^n = [y(1+\epsilon R)]/[(1-\epsilon)(2-y)] \cdot [A_{\parallel} + \tan(\theta/2)A_{\perp}]$, where $y = (E-E')/E$ in the laboratory frame. The neutron ratio was extracted from our $^3$He data in the same way as $A_1^n$.
\begin{table}[htp]
\begin{center}
\caption{$A_1^n$ and $g_1^n/F_1^n$ results.}
\label{tab:results_a1}
\begin{tabular}{lrr}
\multicolumn{1}{c}{$\langle x \rangle$} & \multicolumn{1}{c}{$A_1^n \pm \textrm{stat} \pm \textrm{syst}$} & \multicolumn{1}{c}{$g_1^n/F_1^n \pm \textrm{stat} \pm \textrm{syst}$}  \\
\hline
0.277 & $0.043 \pm 0.060 \pm 0.021$ & $0.044 \pm 0.058 \pm 0.012$ \\
0.325 & $-0.004 \pm 0.035 \pm 0.009$ & $-0.002 \pm 0.033 \pm 0.009$ \\
0.374 & $0.078 \pm 0.029 \pm 0.012$ & $0.053 \pm 0.028 \pm 0.010$ \\
0.424 & $-0.056 \pm 0.032 \pm 0.013$ & $-0.060 \pm 0.030 \pm 0.012$ \\
0.474 & $-0.045 \pm 0.040 \pm 0.016$ & $-0.053 \pm 0.037 \pm 0.015$ \\
0.548 & $0.116 \pm 0.072 \pm 0.021$ & $0.110 \pm 0.067 \pm 0.019$ \\
\end{tabular}
\end{center}
\label{default}
\end{table}%

\begin{figure}[tbp]
\begin{center}
        \includegraphics[width=\columnwidth]{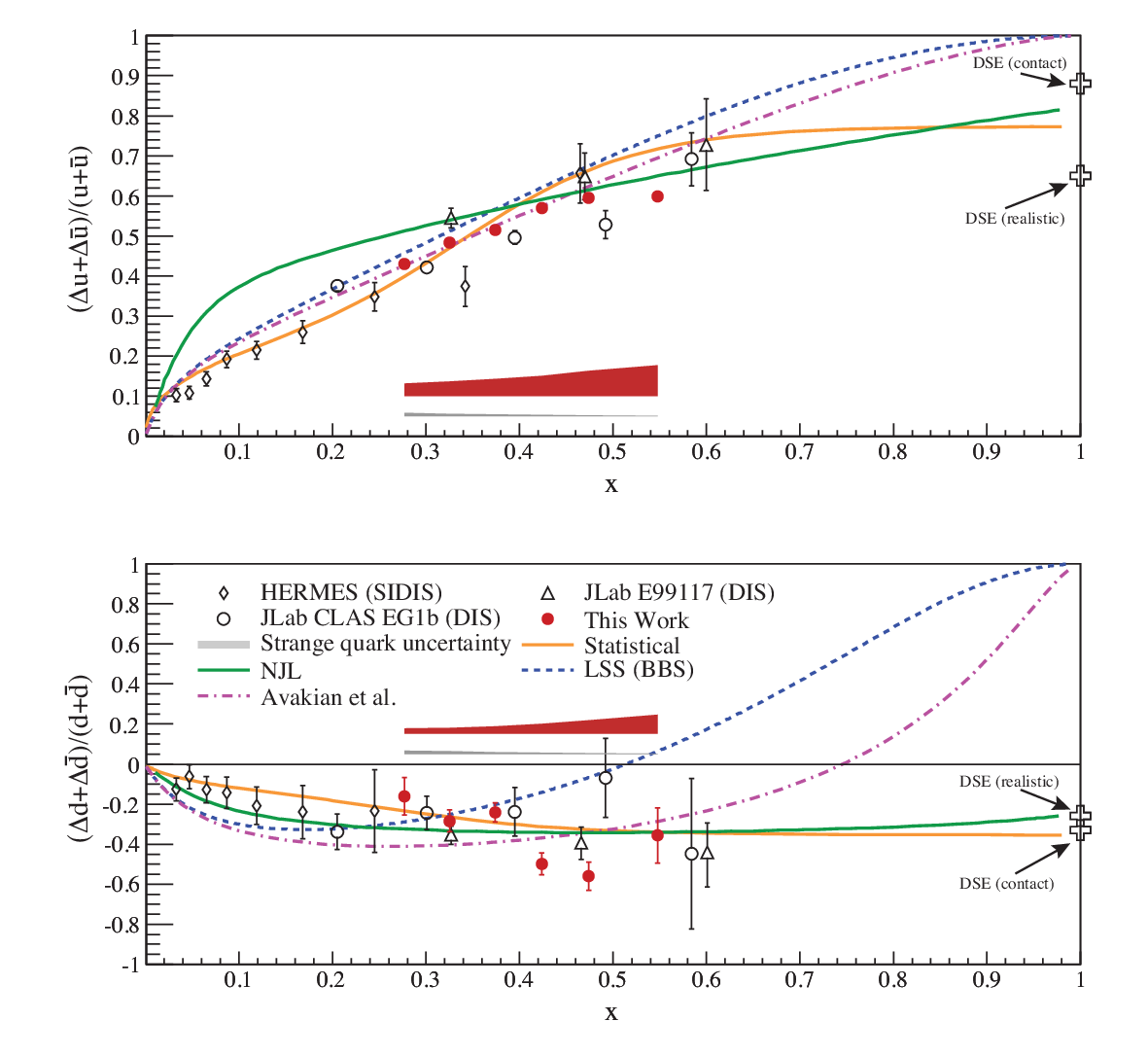}
        \caption{(Color online) Our results (filled circles) for $(\Delta u + \Delta \bar{u})/(u + \bar{u})$ (top, dominated by proton measurements and shown here for reference) and $(\Delta d + \Delta \bar{d})/(d + \bar{d})$ (bottom). The error bars on our results reflect the statistical uncertainties. The upper bands show the total systematic uncertainties on our results, while the lower bands represent the portions of those uncertainties that arise from neglecting the strange-quark contribution. Also plotted are existing semi-inclusive DIS data (HERMES~\cite{ackerstaff:1999}), inclusive DIS data (Jefferson Lab E99117~\cite{zheng:2004} and Jefferson Lab CLAS EG1b~\cite{Dharmawardane:CLASA1meas06}), and models and parameterizations as described in Fig.~\ref{fig:a1n}. More recent semi-inclusive DIS data from HERMES~\cite{Airapetian:PDFsHERMES05} cannot be shown in this figure as the quark and antiquark contributions are separated.}
        \label{fig:flavor_sep}
\end{center}
\end{figure}

Combining the neutron $g_1/F_1$ data with measurements on the proton allows a flavor decomposition to separate the polarized-to-unpolarized-PDF ratios for up and down quarks, giving greater sensitivity than $A_1^n$ to the differences between various theoretical models. When the strangeness content of the nucleon is neglected, these ratios can be extracted at leading order as
\begin{equation}
\frac{\Delta u + \Delta \bar{u}}{u + \bar{u}} = 
\frac{4}{15} \frac{g_1^p}{F_1^p} \left( 4 + R^{du} \right)
- \frac{1}{15} \frac{g_1^n}{F_1^n} \left( 1 + 4 R^{du} \right)
\label{eq:u_flavordecomposition}
\end{equation}
\begin{equation}
\frac{\Delta d + \Delta \bar{d}}{d + \bar{d}} =
\frac{-1}{15} \frac{g_1^p}{F_1^p} \left( 1 + \frac{4}{R^{du}} \right)
+ \frac{4}{15} \frac{g_1^n}{F_1^n} \left( 4 + \frac{1}{R^{du}} \right)
\label{eq:d_flavordecomposition}
\end{equation}
\noindent where $R^{du} \equiv (d + \bar{d})/(u + \bar{u})$ and is taken from the CJ12 parameterization~\cite{cj12}; $g_1^p/F_1^p$ was modeled with world data~\cite{ackerstaff:1997, Abe:e143_98, anthony:1999, Dharmawardane:CLASA1meas06, prok:2014} in the same way as $A_1^p$. An uncertainty of $< 0.009$ for $(\Delta u + \Delta \bar{u})/(u + \bar{u})$ and $<0.02$ for $(\Delta d + \Delta \bar{d})/(d + \bar{d})$ was attributed to the neglect of the strangeness contribution. Our results are given in Table~\ref{tab:results_flavorsep}, and plotted in Fig.~\ref{fig:flavor_sep} along with previous world DIS data and selected model predictions and parameterizations. The $(\Delta u + \Delta \bar{u})/(u + \bar{u})$ results, shown here for reference, are dominated by proton measurements.

\begin{table}[htp]
\begin{center}
\caption{$(\Delta u + \Delta \bar{u})/(u + \bar{u})$ and $(\Delta d + \Delta \bar{d})/(d + \bar{d})$ results. The reported systematic uncertainties include those due to neglecting the strangeness contribution.}
\label{tab:results_flavorsep}
\begin{tabular}{ccc}
$\langle x \rangle$ & $\frac{\Delta u + \Delta \bar{u}}{u + \bar{u}} \pm \delta_{\textrm{stat}} \pm \delta_{\textrm{syst}}$ & $\frac{\Delta d + \Delta \bar{d}}{d + \bar{d}} \pm \delta_{\textrm{stat}} \pm \delta_{\textrm{syst}}$ \\
\hline
0.277 &  $0.430 \pm  0.011 \pm  0.031$ &  $-0.160 \pm  0.094 \pm  0.028$ \\
0.325 &  $0.484 \pm  0.006 \pm  0.037$ &  $-0.283 \pm  0.055 \pm  0.032$ \\
0.374 &  $0.515 \pm  0.005 \pm  0.044$ &  $-0.241 \pm  0.048 \pm  0.039$ \\
0.424 &  $0.569 \pm  0.005 \pm  0.051$ &  $-0.499 \pm  0.054 \pm  0.051$ \\
0.474 &  $0.595 \pm  0.006 \pm  0.063$ &  $-0.559 \pm  0.070 \pm  0.070$ \\
0.548 &  $0.598 \pm  0.009 \pm  0.077$ &  $-0.356 \pm  0.138 \pm  0.097$ \\
\end{tabular}
\end{center}
\label{default}
\end{table}%

Our results for $A_1^n$ and $(\Delta d + \Delta \bar{d})/(d + \bar{d})$ support previous measurements in the range $0.277 \leq x \leq 0.548$. The $A_1^n$ data are consistent with a zero crossing between $x=0.4$ and $x=0.55$, as reported by the Jefferson Lab E99-117 measurement~\cite{zheng:2004a}; extending the original LSS(BBS) pQCD parameterization~\cite{Leader:LSSBBS} to explicitly include quark OAM~\cite{Avakian:OAM07} gives a visibly better match to our data at large $x$. Our leading-order extraction of $(\Delta d + \Delta \bar{d})/(d + \bar{d})$ shows no evidence of a transition to a positive slope, as is eventually required by hadron helicity conservation, in the $x$ range probed. It is not yet possible to definitively distinguish between modern models -- pQCD, statistical, NJL, or DSE -- in the world data to date, but our data points will help constrain further work in the high-$x$ regime. Our results were obtained with a new measurement technique, relying on an open-geometry spectrometer deployed at a large scattering angle with a gas \v{C}erenkov detector to limit the charged-pion background.

Two dedicated DIS $A_1^n$ experiments~\cite{A1_HallA_Proposal, A1_HallC_Proposal} have been approved to run at Jefferson Lab in the coming years, pushing to higher $x$ and studying the $Q^2$ evolution of the asymmetry; one will use an open-geometry spectrometer~\cite{A1_HallA_Proposal}. Our data, in combination with previous measurements, suggest that additional neutron DIS measurements in the region $0.5 \leq x \leq 0.8$ will be of particular interest in establishing the high-$x$ behavior of the nucleon spin structure; in addition, an extension of the DSE-based approach~\cite{roberts:2013} to $x<1$ would be valuable. It is our hope that our data will inspire further theoretical work in the high-$x$ DIS region. 

We gratefully acknowledge the outstanding support of the Jefferson Lab Accelerator Division and Hall A staff in bringing this experiment to a successful conclusion. This material is based upon work supported by the U.S. Department of Energy, Office of Science, Office of Nuclear Physics, under Award Numbers DE-FG02-87ER40315 and DE-FG02-94ER40844 and Contract DE-AC05-06OR23177, under which the Jefferson Science Associates, LLC, operate Jefferson Lab.


\end{document}